\documentclass[9pt,twocolumn,twoside]{osajnl}

\journal{ol} 

\setboolean{shortarticle}{true}

\usepackage{lineno}

\title{Monolithically Integrated Wavelength-meter in InP with measurement bandwidth of 100nm centered on the C band.}

\author[1,*]{Andrea Volpini}
\author[1]{Damiano Massella}
\author[1]{David Alvarez-Outerelo}
\author[2]{Francisco Soares}
\author[1,3]{Francisco J.Diaz-otero}

\affil[1]{University of Vigo, El Telecomunication - Campus Universitario As Lagoas, 36310 Vigo, Spain}
\affil[2]{Soares Photonics, Lisbon, Portugal.}
\affil[3]{AtlanTTic research center, El Telecomunication - Campus Universitario As Lagoas, 36310 Vigo, Spain}

\affil[*]{Corresponding author: andrea.volpini@uvigo.es}

\begin{abstract} 
    In this paper we will explore the creation of a monolithically integrated wavelength meter in InP. This type of devices are a key requirement for many applications and it is especially important to have them integrated with active components like lasers and gain sections.
    We present a wavelength meter based on multiple ring resonators that has been realized in a commercial MPW run and tested using a tunable laser.\newline
    The designed circuit is theoretically capable of resolution down to 1.6pm and a measurement speed down to 500ps within a wavelength range of 100nm.
    
\end{abstract}

\setboolean{displaycopyright}{true}

\begin{document}

\maketitle

\section{Introduction}

The determination of light wavelength is not limited to telecommunication application, but it has also various applications in Wavelength Division Multiplexing, spectroscopy, tunable lasers control and metrology \cite{xiang_integrated_2016,oldenbeuving_high_2013,Komljenovic17widely}.\newline
Among all integrated photonics platforms, Indium Phosphide (InP) is the only platform that allows monolithic integration of lasers and optical gain sections\cite{smit2012generic}.
It is important to have wavelength meters that can be co-integrated with lasers, since this allows better control of integrated tunable laser with all the benefits of integration: lower footprint, better performance and easier fabrication.\newline
InP platforms have a second advantage over the others as they enable higher operation speeds, thanks to high speed modulators, and give future prospects of integration into a more advanced compact photonic system.

A variety of different designs for integrated wavelength meters has been reported in the literature \cite{oldenbeuving_high_2013,dey_demonstration_2013,Taballione17Temperature}. All these works are focused on Silicon or Silicon Nitride platform employing different techniques between ring resonators or differential transmission thought multimode interferometers. The most remarkable result so far have has been able to achieve good performance both in range ($100 nm$) and accuracy ($15 pm$) \cite{Wang17passive}. \newline 
Respect to Silicon and Silicon Nitride platform, InP suffers lower performances since passive components have higher losses and higher bend radius are used in the platforms. This makes particularly hard to design ring resonators with high Free Spectral Range, that are at the hearth of wavelength meters design. \newline
So far the wider bandwidth reported in an InP platform is of 3nm \cite{broeke2013design}.

The design that we present in this paper is specifically thought to mitigate the limitation of the InP platform by use of different ring resonators.\newline
In our design, in order to achieve a high operation bandwidth we employ four microring resonators, each of them contains a phase modulator that can achieve $2\pi$ phase shift and the through port of the ring is connected to an independent photodiode (PD).
The different length of the rings leads to different Free Spectral Range (FSR) that can be used to increase the operational bandwidth of the system. 
The setup work with the modulation of spectral position of ring resonances while monitoring the PDs output. 
The subsequent wavelength measurement is obtained by checking at what modulator bias we observe a minimum in ring transmission. This information creates a set of 4 values that will then be compared with a lookup table, obtained during calibration, to estimate the light wavelength.\newline

The presented design has been fabricated on the generic Indium Phosphide (InP) platform of the Franhofer Heinrich Hertz Institute(HHI), the experimental results in this letter have been obtained with a custom photonic setup in the University of Vigo. \newline

\begin{figure}[t]
   \centering
    \includegraphics[width=0.6\linewidth]{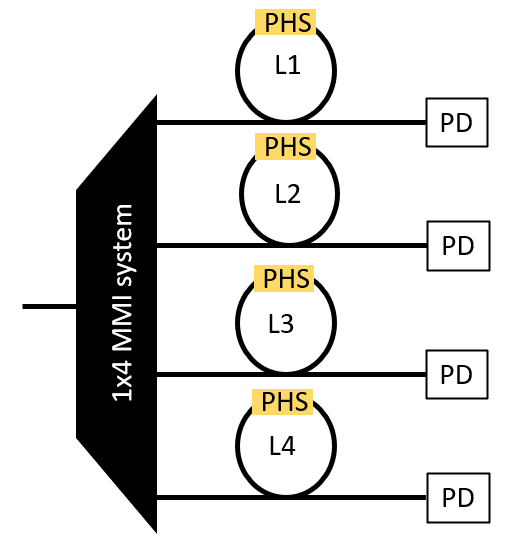}
    \caption{Schematic of the circuit, an unknown monochromatic source is routed to four rings of different length using a combination of MMI splitters. Phase modulators (yellow box) are used to tune the rings on resonance with the source. Transmission spectra are recorded with photodetectors (PD).}
    \label{fig:schematic}
\end{figure}

\section{Methods}

In Figure \ref{fig:schematic} the schematic of the circuit is shown. The device works by routing a monochromatic source to four micro-ring resonators. Each ring contains a phase shifter (PHS) that allows to change the ring resonance when electrically controlled.
Please notice that no "drop" port is present along the rings. In such way unnecessary losses are avoided allowing for higher cavity Q factor. \newline
Calling $\theta(I)$ the phase added by the PHS, the subset of wavelength on resonance with the ring are defined by the expression: 
\begin{equation}\label{eq:res}
\lambda(k) =\dfrac{L \cdot n_{eff}}{k+\dfrac{\theta(V)}{2\pi}} 
\end{equation}

where $k \in \mathbb{N}$, $L$ is the ring length, $n_{eff}$ the effective refractive index and $V$ the voltage applied to the phase shifter.

As mentioned before, the phase $\theta(V)$ can be changed by controlling the value of the applied voltage $V$. If the phase shifter is long enough, $\theta(V)$ can be made greater than $2\pi$ and the ring can be tuned on resonance with any input wavelength. The resonance condition can be easily identified since, when reached, minimum power is detected at the photo detector.\newline  
Equation \ref{eq:res} does not define a bijective relation between the voltage $V$ and a wavelength $\lambda$. In other words, any ring taken individually can operate as a wavelength meter only if its free spectral range ($FSR$) is wider than the band of the unknown source. This problem can be overcome by using carefully designed rings of different length. This is because rings of different lengths have different possible sets of resonances associated with phase shifter voltage $V$. The intersection of such sets determines the input wavelength. To obtain such sets of wavelengths the system is operated as following.

The source to be measured is coupled in the device, and transmitted intensity is measured at the detectors. PHSs are driven to sweep the phase $\theta(V)$ in the interval $[0,2\pi]$, in this way all PDs will have at least a minimum in transmission (Figure \ref{fig:data}) that can be associated to a driving voltage $V$ characteristic of each ring. From each individual resonant voltage the subset of possible resonance wavelengths is deduced.

It is worth to point out that to predict exactly such sets of wavelength a detailed model of: group index dispersion, phase control, temperature dependence and fabrication defects would be necessary. Given the challenge of the task, is much easier to proceed with an external calibration of the system in order to create a look up table where wavelengths are associated to specific voltages .

\section{Ring design}
Rings must be designed of different length in order to allow for bandwidth expansion. It is advantageous to start with the design of  ring $L_1$, that has to be short as possible. Infact, the shorter the ring, the smaller the propagation loss, thus leading to an high $Q$ factor and consequently a small full width half maximum (FWHM) of the resonance. This is of critical importance since FWHM poses a lower limit on the  resolution achievable with the device. Shorter rings also have a wider FSR, that makes the task of expanding the detectable bandwidth easier. 

In general, the  minimum achievable length  for a ring is fixed by a number of design constrains. Every ring must have: a directional coupler, with a proper length to guarantee coupling, a PHS, long enough to guarantee at least a $2\pi$ phase shift, and two 180 degrees bends with the smallest radius of curvature available on the platform. 

The shortest ring we have designed in the HHI platform has a length of $l=3318\mu m$. We call this ring $L_{1}$. When designing the ring, we were afraid of not being able to reach phase shifts up to $2\pi$. We therefore used a PHS of $800\mu m$ length. As it can be seen from figure \ref{fig:data}, we could have been less conservative. A PHS of $300\mu m$ would have worked as well and would have required a voltage of about $5V$ to get  a $2\pi$ phase shift.
\begin{equation}\label{eq:fsr}
FSR=\frac{c}{n_g L} 
\end{equation}

\begin{figure} [h]   
    \centering
        \includegraphics[width=\linewidth]{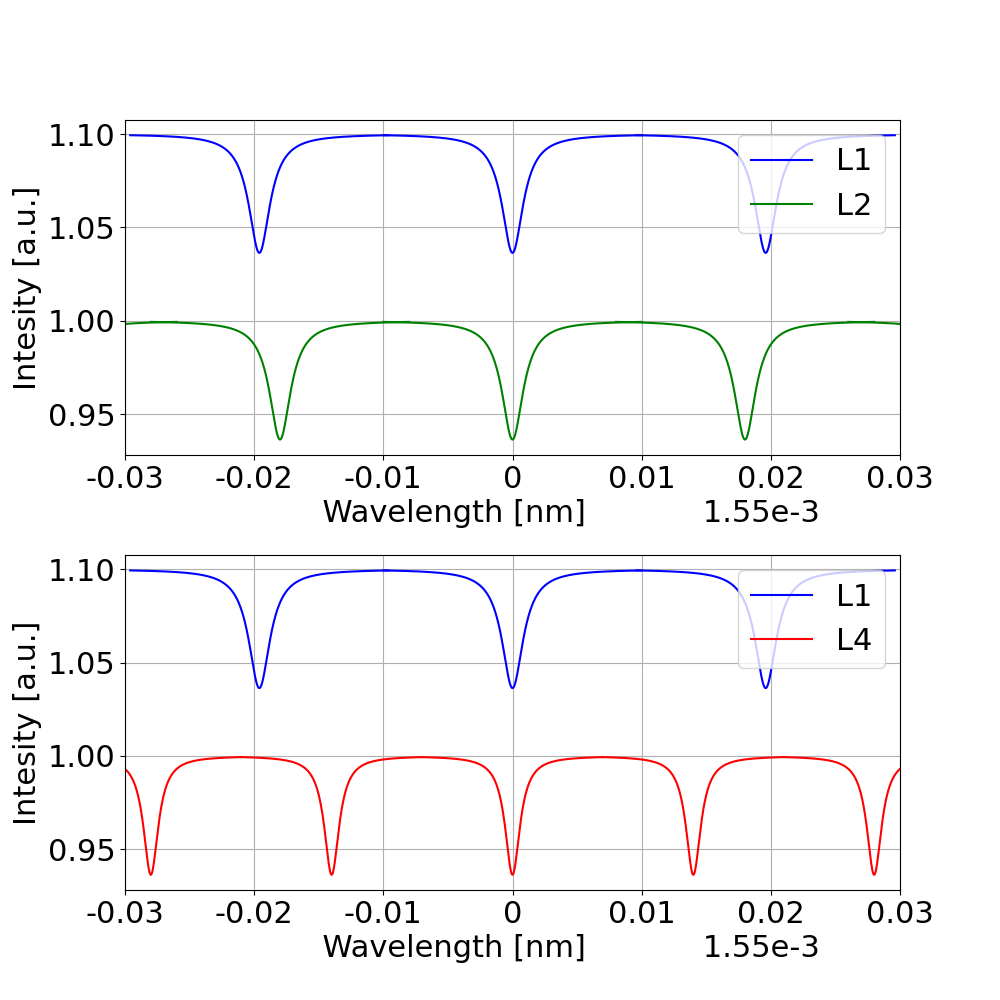}
        \caption{Top figure:
        Ring L1 and Ring L2 have similar FSR and consequently it is not possible to resolve the resonance unambiguously in the interval.\newline
        Bottom figure:Ring L1 and Ring L3 have different spectral range resulting in the possibility of determine the wavelength without ambiguity in the selected interval.}
        \label{fig:trial}
\end{figure}

For ring $L1$, assuming an effective index of $n_eff=3.5$, using equation \ref{eq:fsr} we expect FSR close to $25$ GHz ($0.2nm$) at a central frequency of $193 THz$. With a process that resembles the vernier effect is possible to expand the operational bandwidth of the device. 

Equation \ref{eq:3FSR} \cite{Komljenovic2017} predicts the combined FSR of two rings of length $l_1$, $l_2$. We can verify that a length $L_2=3324 \mu m$ corresponds to  $FSR_{1-2}=14.3THz$ ($100 nm$).

\begin{equation}\label{eq:3FSR}
FSR_{1-2}=FSR_{1}\frac{L_1}{L_1-L_2}
\end{equation}
\newline

 However, this is not sufficient to determine the unknown wavelength over such broad spectrum. With the help of figure \ref{fig:trial}, we can understand better the problem. In the upper plot the the spectra of ring $L_1$ and $L_2$ are simulated. 
 
 Because of the finite $FWHM$ of the resonances we cannot determine if the resonance is at $1550.00 nm$, $1550.02 nm$ or $1449.98 nm$.
 In the lower plot instead we see what happens when we compare compare the spectra of $L_1$ and $L_4$. $L_4 $ as a length of  $l_4=3578 \mu m$ such that  $FSR_1-FSR_4>FWHM$. The degeneracy has been removed and we can state the resonance common to both rings is at $1550.00 nm$. But why don't we use only the rings $L_1$ and $L_4$ then? Because in such case the new combined $FSR$ would be of only $3.318 nm $.
 According to our simulations, using only three rings is not sufficient to cover accurately a $100 nm$ wide band since there are regions in the spectra where a single wavelength cannot be identified. To finally solve the problem we have added ring $L_3$ with $l_3=3578 \mu m$.

\section{Calibration procedure}
\begin{figure}
    \centering
    
      \includegraphics[width=\linewidth]{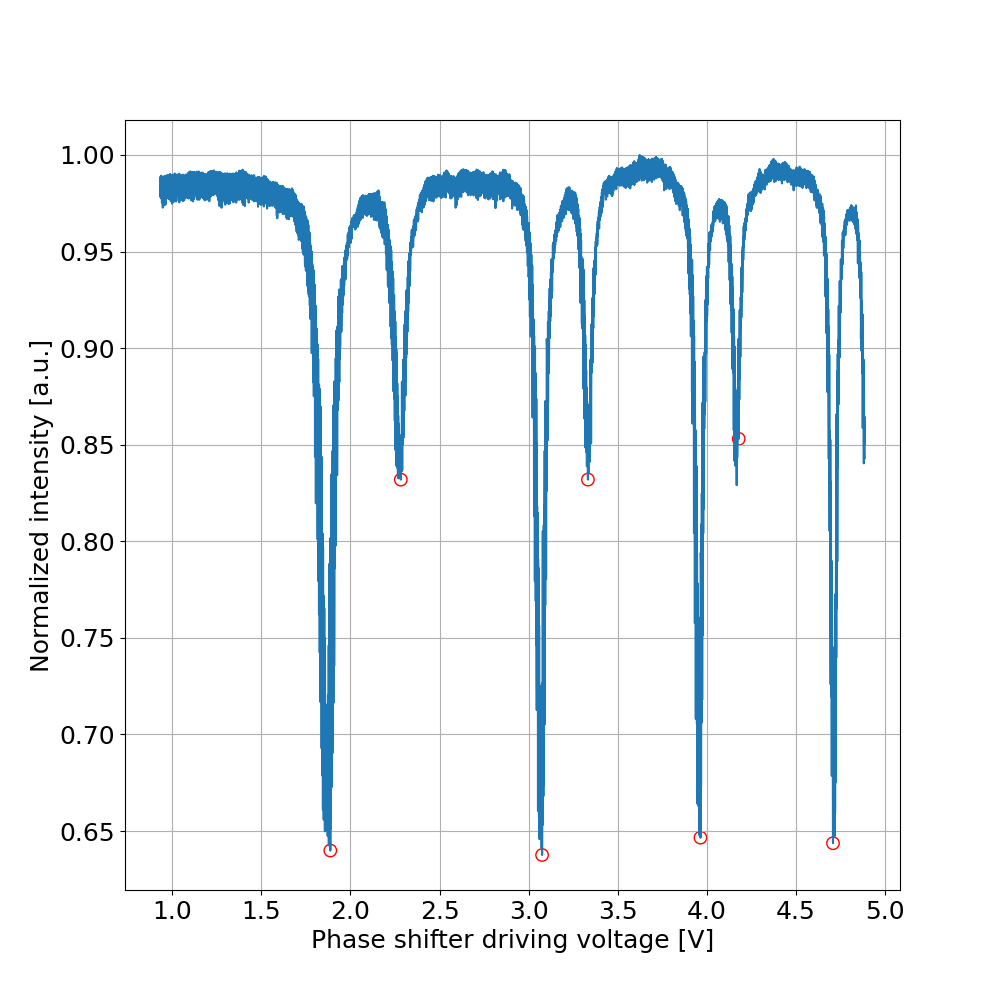}
        \caption{Measured response of ring L1 at a wavelength of 1549.96. The figure is obtained sweeping voltages between $0V$ and $5V$ and registering the power at the integrated photodiode. Using a peak detection alghorithms we can identify the voltages associated with minima, noticing two series of peak at different depth we have decided to consider only odd peaks in our subsequent analysis.}
        \label{fig:data}
\end{figure}

At the moment, we have tested the concept of bandwidth expansion  over the span of $1 nm$ at $1550 nm$. For such limited bandwidth rings $L_1$ and $L_4$ are sufficient to determine the wavelength without ambiguity.
For that we have used a laser source of known wavelength, and acquired the PD response in steps of $20 nm$. In Figure \ref{fig:data} we see an acquisition of the shortest ring at $1549.96 nm$. We see numerous peaks, that implies that we overestimated the necessary PHS minimum length and we could have used a shorter PHS. The peaks are clearly of two types: in particular we notice the odd peaks are much deeper than the even one.
This is likely caused by the propagation of two different optical modes, likely fundamental $TE$ and $TM$. For our calibration, we have decided to consider only the first three odd peaks, that are clearly distinguishable in all our acquisitions.
For any wavelength step we have stored the voltages corresponding to such peaks. We did this both for peaks of ring $L_1$ and $L_4$.

\begin{figure}    
    \centering
        \includegraphics[width=\linewidth]{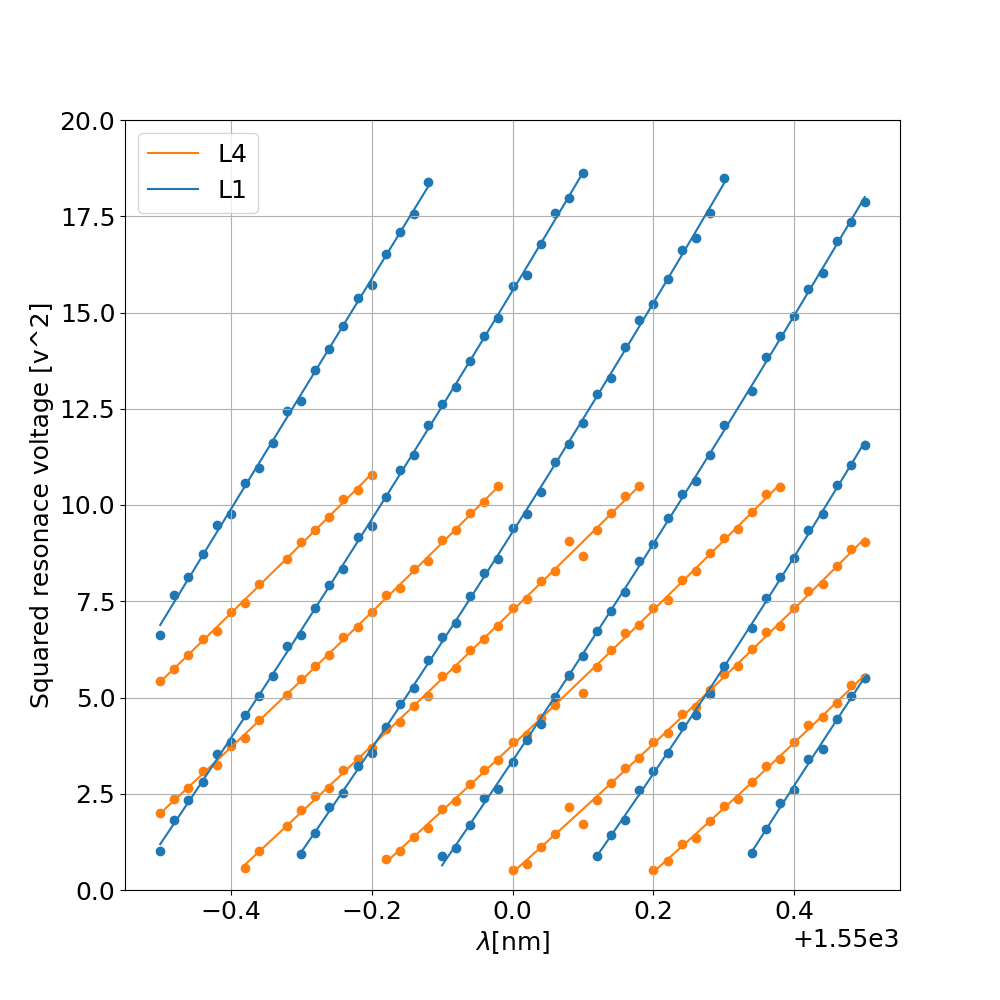}
        \caption{Peak voltages of Ring L1 and Ring L4 respect to the injection wavelength, each point in the graph correspond to a minima: red point are minima of Ring L1, blue points are minima of Ring L4. The data are fitted with a quadratic function in order to create a lookup table.}
        \label{fig:cal}
\end{figure}

\begin{figure}[]
  \centering
         \includegraphics[width=\linewidth]{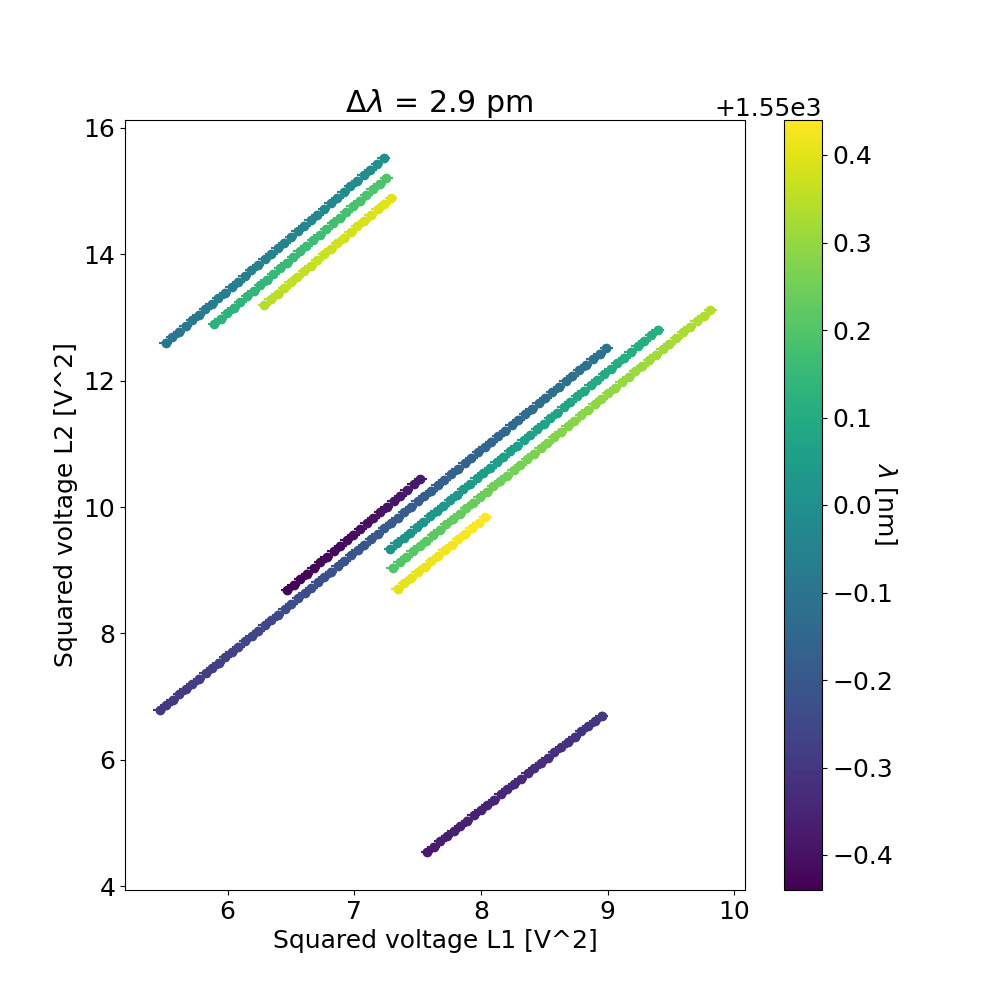}
        \caption{A visualization of the look up table, On the axes the voltages obtained by fitting represented in Fig \ref{fig:cal} for the two different ring. The resulting wavelength is color coded, point are separate by  $4.4 pm$.}
        \label{fig:final}
\end{figure}

In Fig. \ref{fig:cal} we have plotted the square of this voltage versus the corresponding wavelength. We plot the square because PHs are thermally controlled, and the electric power dissipated is proportional to the voltage squared. Every single group of resonances is fitted with a linear function. Using this fit is possible to create the look up table that associates the two voltages ($V_1$,$V_2$) to the corresponding wavelength.
In the phase diagram of Fig. \ref{fig:final} we have all the data displayed in the look up table. The horizontal axis shows the squared voltage applicable to the ring $L_1$, the vertical ones refers to the squared voltage applicable to $L_2$. Points in the phase space are identified by couples of values ($V_1^2$,$V_2^2$). The color of the point codifies the wavelength according to the color bar on the right. Error bars are also displayed. They have been obtained by computing the standard variation of experimental points from the predicting fitting model. We see that such error-bars correspond to an uncertainty of $\pm 1.5 pm$ on the estimated wavelength. To operate the wavelength meter, a script where experimental ($V_1,V_2$) are associated to the corresponding wavelength is necessary. Despite the extension of the error bars, the identification of an unique wavelength as been always possible in the tested range of $1nm$.

\section{Reading speed limit} 
We have realized that there is a minimum speed limit necessary to estimate the wavelength. Such limit is linked to light propagation into ta ring. Indeed,a ring resonator has a typical cavity lifetime that is determined by the $Q$ factor: $Q=\omega\tau_s$ . A time $T=3\tau_s$ is necessary before considering the ring to be stabilized\cite{Biasi_2019}. To cover the full FSR of any of the rings we need a number of steps in the voltage applied to the PHS. 
We chose this number of steps ($m$), in order not to lose resolution but also not to oversample the PD response. Given that the limit resolution is equal to the $FWHM$ of the ring:
 \begin{equation}\label{eq:voltage_step}
 m=\frac{2FSR}{FWHM}=\frac{2FSR \times Q}{\omega}
 \end{equation}
We can express this equation in therms of $Q$ to estimate the time $T$ necessary for a complete measurement as:
 
 \begin{equation}\label{eq:measurement_time}
 T=m\tau_s=\frac{6FSR\times Q^2}{\omega^2}
 \end{equation}
 We can notice from this equation that the measurement speed scales with the square of the $Q$ factor.
 In the following table we estimate the speed of our device for different Q factors at constant single ring $FSR=30$ GHz and $\omega=193$ THz.
 
 \begin{table}
 \centering
 \caption{Table representing the correlation between the Quality factor of the cavity, minimum measurement Time and measurement resolution. The results are obtained applying equation \ref{eq:measurement_time}.}
 \label{tab:speed}
 \begin{tabular}{c  c  c} 
 \hline
 Q factor &  T & Resolution \\ 
 \hline
 $10^4$ & $500 ps$ & $160  pm $ \\ 
 
 $10^5$ & $50 ns$ &  $20  pm $\\
 
 $10^6$ & $5 \mu s$  & $1.6  pm $\\
\end{tabular}
\end{table}
An important remark is necessary, the measurement speed does not depend on the number of rings since all the rings are measured simultaneously. 


\section{Conclusions}

 A photonic-integrated wavelength-meter based on multiple ring resonators has been realized in InP. The correct operation of the device has been verified for a limited bandwidth of $1nm$. New measures are expected to be carried out with a much broader band in order to verify the stability and reliability over a span of $100 nm$. Moreover, upon realization of wire bonding connections, it will be possible to test the maximum readout speed of the device.


\section{Acknowledgements}

Project developed in the framework of the 
European Doctorate in Indium Phosphide PIC Fabrication Technology (EDIFY) project. This work has received financial support from MSCA H2020-ITN-2018-EDIFY (Contract number 813467)
\section{Disclosures}
The authors declare no conflicts of interest.

\bibliography{30_references/references_new}

\begin{thebibliography}{10}
\newcommand{\enquote}[1]{``#1''}

\bibitem{xiang_integrated_2016}
C.~Xiang, M.~A. Tran, T.~Komljenovic, J.~Hulme, M.~Davenport, D.~Baney,
  B.~Szafraniec, and J.~E. Bowers, {\protect\JournalTitle{Opt. Lett.}}
  \textbf{41}, 3309 (2016).

\bibitem{oldenbeuving_high_2013}
R.~M. Oldenbeuving, H.~Song, G.~Schitter, M.~Verhaegen, E.~J. Klein, C.~J. Lee,
  H.~L. Offerhaus, and K.-J. Boller, {\protect\JournalTitle{Opt. Express}}
  \textbf{21}, 17042 (2013).

\bibitem{Komljenovic17widely}
T.~Komljenovic, L.~Liang, R.-L. Chao, J.~Hulme, S.~Srinivasan, M.~Davenport,
  and J.~E.~Bowers, {\protect\JournalTitle{Applied Sciences}} \textbf{7}
  (2017).

\bibitem{smit2012generic}
M.~Smit, X.~Leijtens, E.~Bente, J.~van~der Tol, H.~Ambrosius, D.~Robbins,
  M.~Wale, N.~Grote, and M.~Schell, \enquote{A generic foundry model for
  inp-based photonic ics,} in \emph{Optical Fiber Communication Conference,}
  (Optical Society of America, 2012), pp. OM3E--3.

\bibitem{dey_demonstration_2013}
R.~Dey, J.~Doylend, J.~Ackert, A.~Evans, P.~Jessop, and A.~Knights,
  {\protect\JournalTitle{Opt. Express}} \textbf{21}, 23450 (2013).

\bibitem{Taballione17Temperature}
C.~Taballione, T.~Agbana, G.~Vdovin, M.~Hoekman, L.~Wevers, J.~Kalkman,
  M.~Verhaegen, P.~J. van~der Slot, and K.-J. Boller,
  \enquote{{Temperature-drift-immune wavelength meter based on an integrated
  micro-ring resonator},} in \emph{Integrated Optics: Physics and Simulations
  III,} , vol. 10242 P.~Cheben, J.~Čtyroký, and I.~Molina-Fernández, eds.,
  International Society for Optics and Photonics (SPIE, 2017), pp. 39 -- 48.

\bibitem{Wang17passive}
P.~Wang, A.~M. Hatta, H.~Zhao, W.~Yang, J.~Ren, Y.~Fan, G.~Farrell, and
  G.~Brambilla, {\protect\JournalTitle{Opt. Express}} \textbf{25}, 2939 (2017).

\bibitem{broeke2013design}
R.~Broeke, D.~Melati, F.~Morichetti \emph{et~al.}, \enquote{Design and
  performance of a packaged inp wavelength meter,} in \emph{18th Annual
  Symposium of the IEEE Photonics Benelux Chapter,}  (2013), pp. 287--290.

\bibitem{Komljenovic2017}
T.~Komljenovic, L.~Liang, R.-L. Chao, J.~Hulme, S.~Srinivasan, M.~Davenport,
  and J.~E.~Bowers, {\protect\JournalTitle{Applied Sciences}} \textbf{7}
  (2017).

\bibitem{Biasi_2019}
S.~Biasi, P.~Guilleme, A.~Volpini, G.~Fontana, and L.~Pavesi,
  {\protect\JournalTitle{Journal of Lightwave Technology}} \textbf{37}, 5091
  (2019).

\end{thebibliography}

\bibliographyfullrefs{30_references/references_new}


\ifthenelse{\equal{\journalref}{aop}}{%
\section*{Author Biographies}
\begingroup
\setlength\intextsep{0pt}
\begin{minipage}[t][6.3cm][t]{1.0\textwidth} 
  \begin{wrapfigure}{L}{0.25\textwidth}
    \includegraphics[width=0.25\textwidth]{john_smith.eps}
  \end{wrapfigure}
  \noindent
  {\bfseries John Smith} received his BSc (Mathematics) in 2000 from The University of Maryland. His research interests include lasers and optics.
\end{minipage}
\begin{minipage}{1.0\textwidth}
  \begin{wrapfigure}{L}{0.25\textwidth}
    \includegraphics[width=0.25\textwidth]{alice_smith.eps}
  \end{wrapfigure}
  \noindent
  {\bfseries Alice Smith} also received her BSc (Mathematics) in 2000 from The University of Maryland. Her research interests also include lasers and optics.
\end{minipage}
\endgroup
}{}

\end{document}